\documentclass[a4paper,12pt]{article}
\usepackage{amsmath}
\usepackage{citehack}
\usepackage{bm}
\usepackage{graphicx}
\usepackage{epstopdf}
\usepackage{color}
\usepackage{comment}
\usepackage{hyperref}
\usepackage{float}
\usepackage{tikz, xcolor, pgfplots}
\usetikzlibrary{shapes, arrows}
\begin{document}

\title{2D microscopic and macroscopic simulation of water and porous material interaction\thanks{topic JINR LIT No. 05-6-1118-2014/2019, protocol No. 4596-6-17/19.}}
\author{E.G. Nikonov$^1$, M.Pavlu\v{s}$^2$, M. Popovi\v{c}ov\'a$^2$\\
\begin{minipage}{10cm}
\begin{center}
\small\em
\vspace{4mm}
$^1$Joint Institute for Nuclear Research,\\ 141980 Dubna, Moscow Region, Russia\\ email: e.nikonov@jinr.ru \\
\vspace{4mm}
$^2$University of Pre\v{s}ov,\\ str. Kon\v{s}tantinova 16, 080 01 Pre\v{s}ov,  Slovakia\\ email: miron.pavlus@unipo.sk, maria.popovicova@unipo.sk
\end{center}
\end{minipage}
} 
\date{}
\maketitle

\begin{abstract}
In various areas of science, technology, environment protection, construction, it is very important to study processes of porous materials interaction with different substances in different aggregation states. From the point of view of ecology and environmental protection it is particularly actual to investigate processes of porous materials interaction with water in liquid and gaseous phases. Since one mole of water contains $6,022140857\cdot 10^{23}$ molecules of $\mathtt{H_2O}$,
macroscopic approaches considering the water vapor as continuum media in the framework of classical aerodynamics are mainly used to describe properties, for example properties of water vapor in the pore. In this paper we construct and use for simulation the macroscopic two-dimensional diffusion model \cite{Bitsadze1980} describing the behavior of water vapor inside the isolated pore. Together with the macroscopic model it is proposed microscopic model of the behavior of water vapor inside the isolated pores. This microscopic model is built within the molecular dynamics approach\cite{Gould}. In the microscopic model a description of each water molecule motion is based on Newton classical mechanics  considering interactions with other molecules and pore walls. Time evolution of water vapor - pore system is explored. Depending on the external to the pore conditions the system evolves to various states of equilibrium, characterized by different values of the macroscopic characteristics such as temperature, density, pressure. Comparisons of results of molecular dynamic simulations with the results of calculations based on the macroscopic diffusion model and experimental data allow to conclude that the combination of macroscopic and microscopic approach could produce more adequate and more accurate description of processes of water vapor interaction with porous materials.
\end{abstract}

\paragraph{Introduction}

Experimental and theoretical investigations of liquid and gas interaction with porous media are very important for many fields of science and technology. An interaction of water in various phase states with construction materials; filtering and separation of different substances in chemical, biological industries and medicine; protection against toxic and radioactive substances in gaseous and droplet forms etc. This is not a complete list of fields where it is very necessary to have an information about characteristics of processes of water interaction with porous media. Porous materials can be organic or inorganic origin. The most frequent porous materials are organic materials, polymeric foams etc. A large number of inorganic porous materials has also been developed, e.g. for insulation, cushioning, impact protection, catalysis, membranes, construction materials etc.

Porous material can be of the following types.
\begin{itemize}
\item Materials with different pore sizes (from nanometer to millimeter)
\item Ordered or irregular arrangement of pores
\item Various chemical compositions (metal, oxides…)
\item Preparated by different way on the base of various preparative approaches.
\end{itemize}

Pores can be distinguishes by the following features: accessibility and shape. Accessibility means that pores can be of the following types: closed pores, open pores, blind pores (also called dead-end or saccate pores) and through pores. Shape means that pores can be of the following types: cylindrical open, cylindrical blind, ink-bottle-shaped, funnel shaped and roughness \cite{Rouquerol1994}.

Moreover, in accordance with the IUPAC\footnote{The International Union of Pure and Applied Chemistry}  classification, pores are characterized by their sizes\cite{McNaught1997}.
\begin{itemize}
\item Micropores (smaller than 2 nm): larger than typical mean free path length of typical fluid. 
\item Mesopores (between 2 and 50 nm):  same order or smaller than the mean free path length. 
\item Macropores (larger than 50 nm):  pore size comparable to the molecules. 
\end{itemize}

All artificial and natural porous material have a very complicated porous structure with pores of very intricate geometrical forms. So if we need to accurately describe processes of water--pore interaction for individual pore we need to numerically solve,  for example, diffusion equation with very complicated boundary conditions. In this case, if we use molecular dynamic for finding of, for example, diffusion coefficient space distribution, we can get more accurate numerical solution for diffusion equations. Evidently this approach is justified if and only if  it is necessary to obtain a description of water interaction with individual pore as accurate as possible.

\begin{figure}[H]
\center{\includegraphics[width=0.5\linewidth]{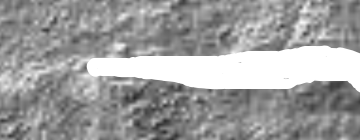}}
\caption{Cross-section of slit-like pore.}
\label{pore_cs}
\end{figure}

In this paper we compare continual macroscopic diffusion model and discrete microscopic molecular dynamic model to investigate differences between these two approaches in description of water vapor interaction with individual pore. 
\paragraph{General problem}
Because of a high level of complexity of water-pore interaction which heavily depends on the form of the pore we consider in our work as the first step of investigation two-dimensional~(2D) model of a slit-like pore with cross-section as shown in figure \ref{pore_cs}.

Further in the paper we compare micro and macro approaches for describing of water vapor interaction with 2D slit-like pore. Micro approach is based on ''ab initio'' molecular dynamic simulations. And macro approach is based on macroscopic diffusion model. The main goal of our investigation is understanding of possibilities to combine the two approaches to achieve higher accuracy in description of processes in real water-pore systems.
\paragraph{Molecular dynamic model}\label{Micro}

Molecular dynamic simulation of microscopic system evolution is based on representation of object modeling as a system of interacting particles (atoms or molecules). The evolution of the system is a result of a motion of the particles mentioned above. Particle coordinates in each subsequent time step are calculated by integration of equations of motion. These equations contain potentials of particle interactions with each other and with an external environment.

In classical molecular dynamics, the behaviour of an individual particle is described by the Newton equations of motion\cite{Gould}, \cite{Shaitan1999}.

In our work for a simulation of particle interaction, we use Lennard-Jones potential\cite{LJ} with parameters $\sigma = 3.17\mbox{\AA}$ and $\varepsilon = 6.74\cdot 10^{-3} eV$. This potential is the most used to describe of the evolution of water in liquid and saturated vapor forms.

One of the important thing for the molecular dynamic simulations to account for the effects of energy interchanging with the ambient is a special algorithm which is called a thermostat. In our work, we use  Berendsen thermostat\cite{Berendsen1984} to account for the effects of heat transfer between a vapor in the pore and the external environment. Coefficient of the velocity recalculation $\lambda(t)$ at every time step $t$ depends on the so called 'rise time' of the thermostat $\tau_B$ which belongs to the interval $ [0.5,2] \ \mbox{ps}$. This thermostat uses alternating nonlinear friction in equations of motion.

Equations of motion were integrated by Velocity Verlet method \cite{Verlet1967}.

\paragraph{Macroscopic diffusion model}\label{Macro}
Let us denote the water vapor concentration as $w_v(x,y,t)$ [$ ng/(nm)^2$] in the point $(x,y,t)$ where $x,y$ are space independent variables and $t$ is time independent variable. Then, we consider the following macroscopic diffusion model
\begin{equation}\label{eq01}
\frac{\partial w_v}{\partial t}=D\Big(\frac{\partial^2 w_v}{\partial x^2}+\frac{\partial^2 w_v}{\partial y^2}\Big)\qquad
0<x<l_x\qquad 0<x<l_y\qquad t>0
\end{equation} 
\begin{equation}\label{eq02}
w_v(x,y,0)=w_{v,0}\qquad 0\leq x\leq l_x\qquad 0\leq x\leq l_y
\end{equation} 
\begin{equation}\label{eq03}
\left.\frac{\partial w_v}{\partial n}(t)\right|_{(x,y)\in\Gamma_2\cup\Gamma_3\cup\Gamma_4}=0\qquad t>0
\end{equation}
 
\begin{equation}\label{eq04}
\left.-D\frac{\partial w_v}{\partial x}(t)\right|_{(l_x,y)\in\Gamma_1}=\beta [w_v(l_x,y,t)-w_{v,out}(t)]\qquad 0\leq x\leq l_y\qquad t>0.
\end{equation} 

\begin{figure}[h]
\begin{center}
\begin{tikzpicture}[scale=1.5]
    \draw [<->,thick] (0,2) node (yaxis) [above] {$y$}
        |- (3.5,0) node (xaxis) [right] {$x$};
    \draw (0,1.5) coordinate (y_1) -- (3,1.5) coordinate (y_2);
    \draw [white,thick,dashed] (3,0) coordinate (x_1) -- (3,1.5) coordinate (x_2);
    \coordinate (c) at (intersection of y_1--y_2 and x_1--x_2);
   \draw[dashed] (yaxis |- c) node[left] {$l_y$}
        -| (xaxis -| c) node[below] {$l_x$};
    \draw[<-] (3.1,1)--(3.5,1.5) node (c1) [right] {$\Gamma_1$}; 
    \draw[<-] (1,1.6)--(1.4,2) node (c2) [right] {$\Gamma_2$}; 
    \draw[->] (-0.5,0.4)--(-0.1,0.8) node (c3) [anchor= north east] {$ \Gamma_3\quad$};
    \draw[->] (1.1,-0.5)--(1.5,-0.1) node (c3) [anchor= north east] {$ \Gamma_4\quad$};
\end{tikzpicture}
\end{center}
\caption{Shape of 2D pore.} \label{fig1}
\end{figure}
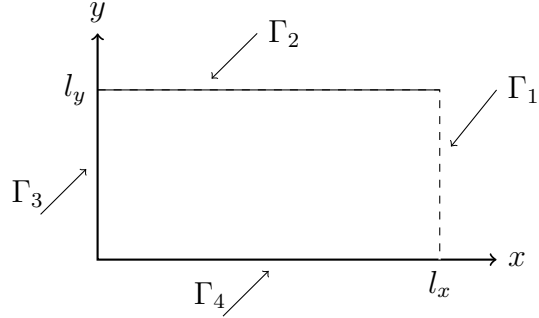
\noindent
where $D$ is the diffusion coefficient [$(nm)^2/ps$]; $l_x, l_y$ are 2D pore dimensions [$nm$]; $w_{v,0}$ is the initial concentration
of water vapor; $\Gamma_1,\Gamma_2,\Gamma_3,\Gamma_4$ are boundaries of 2D pore; $\beta$ is the coefficient of water vapor transfer from pore space to outer space [$nm/ps$]; $w_{v,out}(t)$ is the water vapor concentration in outer space [$ng/(nm)^2$].

We suppose that the outer space water vapor concentration is expressed as
$$
w_{v,out}(t)=\varphi_0\cdot w_{sv}(T_0), 
$$       
where $\varphi_0$ is the relative humidity of outer space ($0\le\varphi_0\le 1$) and $w_{sv}(T_0)$ is saturated water vapor concentration at outer temperature $T_0.$ 

The linear problem (\ref{eq01})--(\ref{eq04}) can be solved exactly by means of the variables separation method 
\cite{Bitsadze1980} 
and the result of the solution is the following 
\begin{equation}\label{eq05}
w_v(x,y,t)=w_{sv}(T_0)\cdot\varphi_0+\Big[w_{v,0}-w_{sv}(T_0)\cdot\varphi_0\Big]\cdot \\ \sum_{m=1}^{\infty}\sum_{n=0}^{\infty}e^{D\lambda_{mn} t}c_{mn}\cos(\alpha_{xm} x)\cos(\alpha_{yn} y) 
\end{equation}
$$
\qquad 0\leq x\leq l_x\qquad 0\leq y\leq l_y\qquad t\ge0.
$$
Here, $c_{mn}$ are coefficients of unity expansion
$$
c_{mn}=
  \begin{cases}
    \frac{4\sin(\alpha_{xm} l_x)}{2 l_x\alpha_{xm}+\sin(2\alpha_{xm} l_x)}       & \quad \text{if } n=0;\ \ m=1,2,3,\dots\\
    0                                                                                                                          & \quad \text{if } n=1,2,3,\dots;\ \ m=1,2,3,\dots\\
  \end{cases}
$$ 
and $\lambda_{mn}$ are eigenvalues where 
$$
\lambda_{mn}=-\alpha_{xm}^2-\alpha_{yn}^2\qquad \alpha_{yn}=\frac{n\pi}{l_y}\qquad n=0,1,2,\dots  
$$
and $\alpha_{xm}$ are solutions of the equations
$$
\alpha_{xm}\cdot\tan(\alpha_{xm} l_x)=\beta/D,\qquad m=1,2,3,\dots. 
$$

\paragraph{Computer simulation of micro-model}\label{microsim}
We consider 2D macropore in the micro model with dimensions $l_x=1\mu m,$ $l_y=1\mu m.$ 
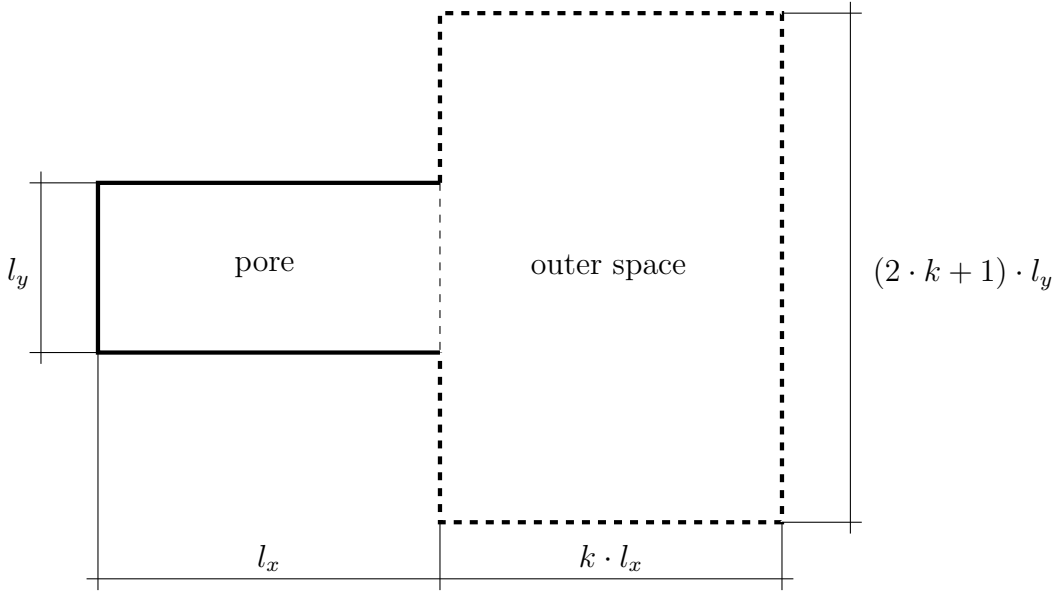
\begin{figure}[h]
\begin{center}
\begin{tikzpicture}[scale=1.5]
    \draw[ultra thick] (3,0)--(0,0)--(0,1.5)--(3,1.5);
    \filldraw[black] (1.1,0.75) node[anchor=west] {pore};
    \draw[dashed] (3,1.5)--(3,0);
    \draw[dashed][ultra thick] (3,1.5)--(3,3)--(6,3)--(6,-1.5)--(3,-1.5)--(3,0); 
    \filldraw[black] (3.7,0.75) node[anchor=west] {outer space};
    \draw (-0.6,0)--(0,0); 
    \draw (-0.6,1.5)--(0,1.5); 
    \draw (-0.5,-0.1)--(-0.5,1.6); 
    \filldraw[black] (-0.5,0.7) node[anchor=east] {$l_y$};  
    \draw (0,0)--(0,-2.1); \draw (3,-1.5)--(3,-2.1); \draw (6,-1.5)--(6,-2.1); \draw (-0.1,-2)--(6.1,-2);
    \filldraw[black] (1.5,-1.6) node[anchor=north] {$l_x$};
    \filldraw[black] (4.5,-1.6) node[anchor=north] {$k\cdot l_x$};
    \draw (6,-1.5)--(6.7,-1.5); \draw (6,3)--(6.7,3); \draw (6.6,-1.6)--(6.6,3.1);  
    \filldraw[black] (6.7,0.7) node[anchor=west] {$(2\cdot k+1)\cdot l_y$};  
\end{tikzpicture}
\end{center}
\caption{2D pore and outer space.} \label{fig2}
\end{figure}
The outer space in this micro model reflects as a space right to the pore, see Fig. \ref{fig2} (dashed line) which size, one can change by means of the parameter $k.$ All sides of the outer space satisfy to the periodic boundary conditions. The left pore side reflects the inner molecules due to the boundary condition (\ref{eq03}) but also provides the periodic boundary conditions for a part of outer space.    

\begin{figure}[H]
\begin{center}
\includegraphics[totalheight=6cm,angle=0,keepaspectratio]{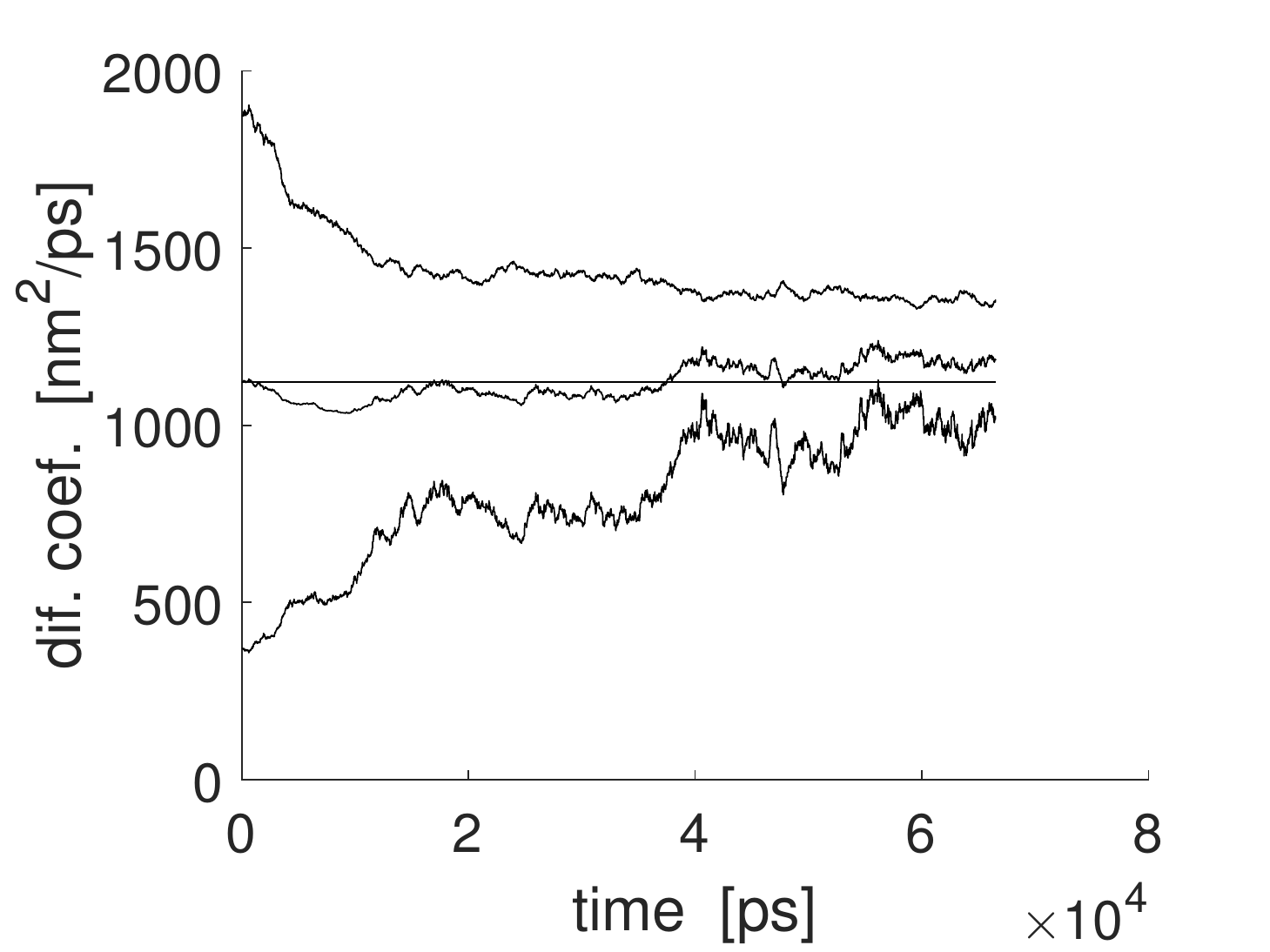}~
\includegraphics[totalheight=6cm,angle=0,keepaspectratio]{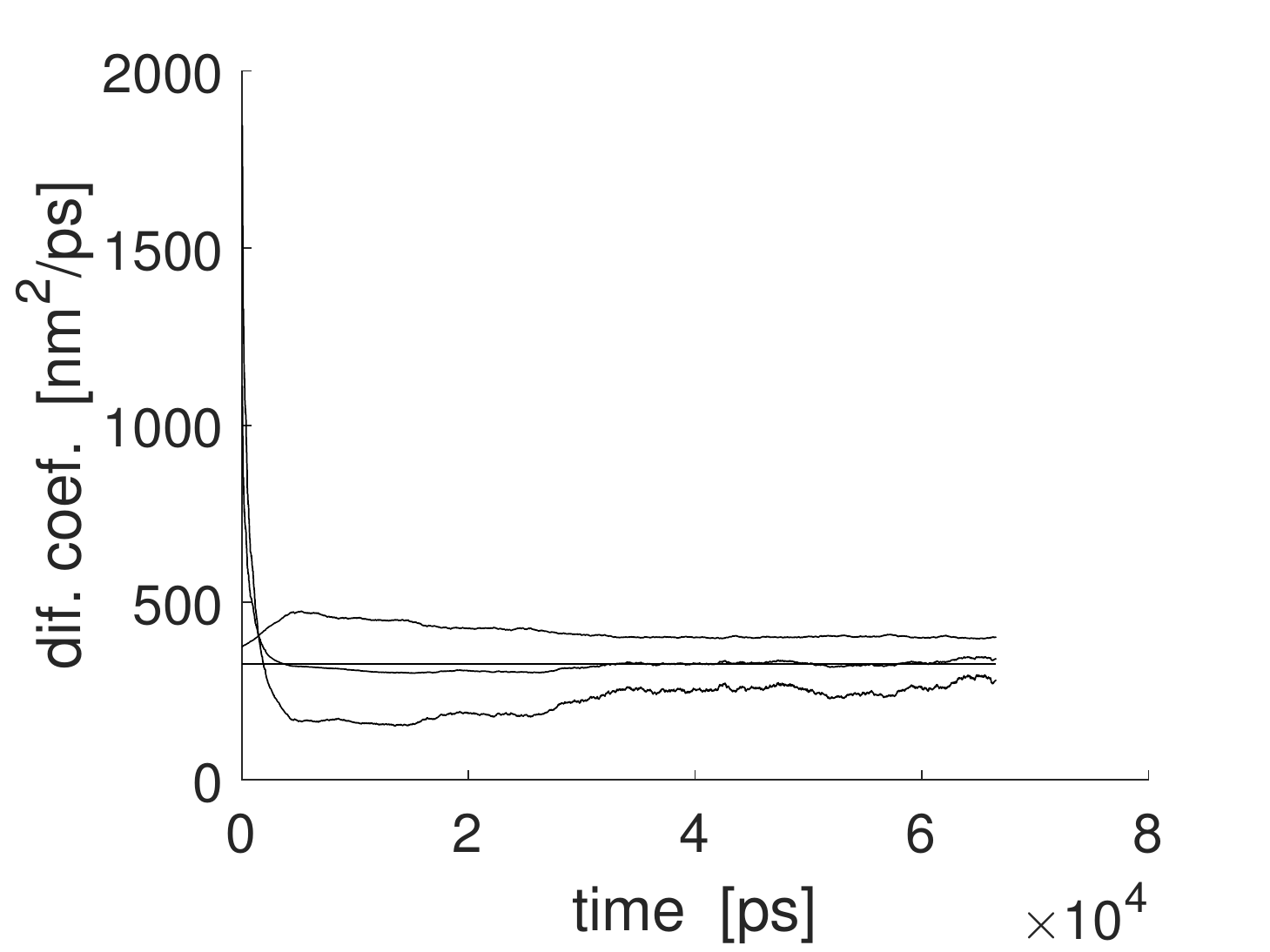}
\end{center}
\caption{Diffusion coefficients for drying process of the pore (left) and for wetting process of the pore (right), final time $t_0=66532$ psec.}\label{Fig_Dif_coef}
\end{figure}

The main characteristic of the diffusion process is the diffusion coefficient. One goal of the micro-model is to construct a constant diffusion coefficient which will then be used in the macro-model. In the figure \ref{Fig_Dif_coef} (left), we show the diffusion coefficients for both the pore $D_p(t)$ (upper predominantly decreasing curve) and the outer space $D_{out}(t)$ (lower predominantly increasing curve) for 244 starting $H_2O$ molecules in the pore. The middle curve is a mean of two previous diffusion coefficients and the constant diffusion coefficient value $D=1121.58$ [m$^2$/sec] is represented in the \ref{Fig_Dif_coef} (left) as a horizontal line that was calculated according to the formula (\ref{eq06})
\begin{equation}\label{eq06}
D=\frac{1}{t_0}\int_0^{t_0}\frac{D_p(t)+D_{out}(t)}{2}dt. 
\end{equation}
The figure \ref{Fig_Dif_coef} (left) corresponds to a drying process of the pore when in the pore, we suppose 244 starting $H_2O$ molecules while in the outer space for $k=2,$ we locate 490 starting $H_2O$ molecules. This means that the pore contains saturated water vapor at temperature $T_0=25$ $^oC$ and at pressure $p_0=3.17$ $kPa$ and the outer space contains only $20~\%$ of it. Thus, the outer space has 20 $\%$ relative humidity.   

The figure \ref{Fig_Dif_coef} (right) corresponds to a wetting process of the pore when in the pore, we locate 49 starting $H_2O$ molecules while in the outer space for $k=2,$ we locate 2440 starting $H_2O$ molecules. This means that the outer space contains saturated water vapor at temperature $T_0=25$ $^oC$ and at pressure $p_0=3.17$ $kPa$ and the pore contains only $20~\%$ of it. Here, the upper curve represents the pore diffusion coefficient, the down curve represents the diffusion coefficient for outer space, the middle curve, as before, is a mean of two previous diffusion coefficients and the constant diffusion coefficient value $D=326.88$ [m$^2$/sec] calculated according the formula (\ref{eq06}) is represented in the Fig. \ref{Fig_Dif_coef} (right) as a horizontal line.     

\begin{figure}[H]
\begin{center}
\includegraphics[totalheight=6cm,angle=0,keepaspectratio]{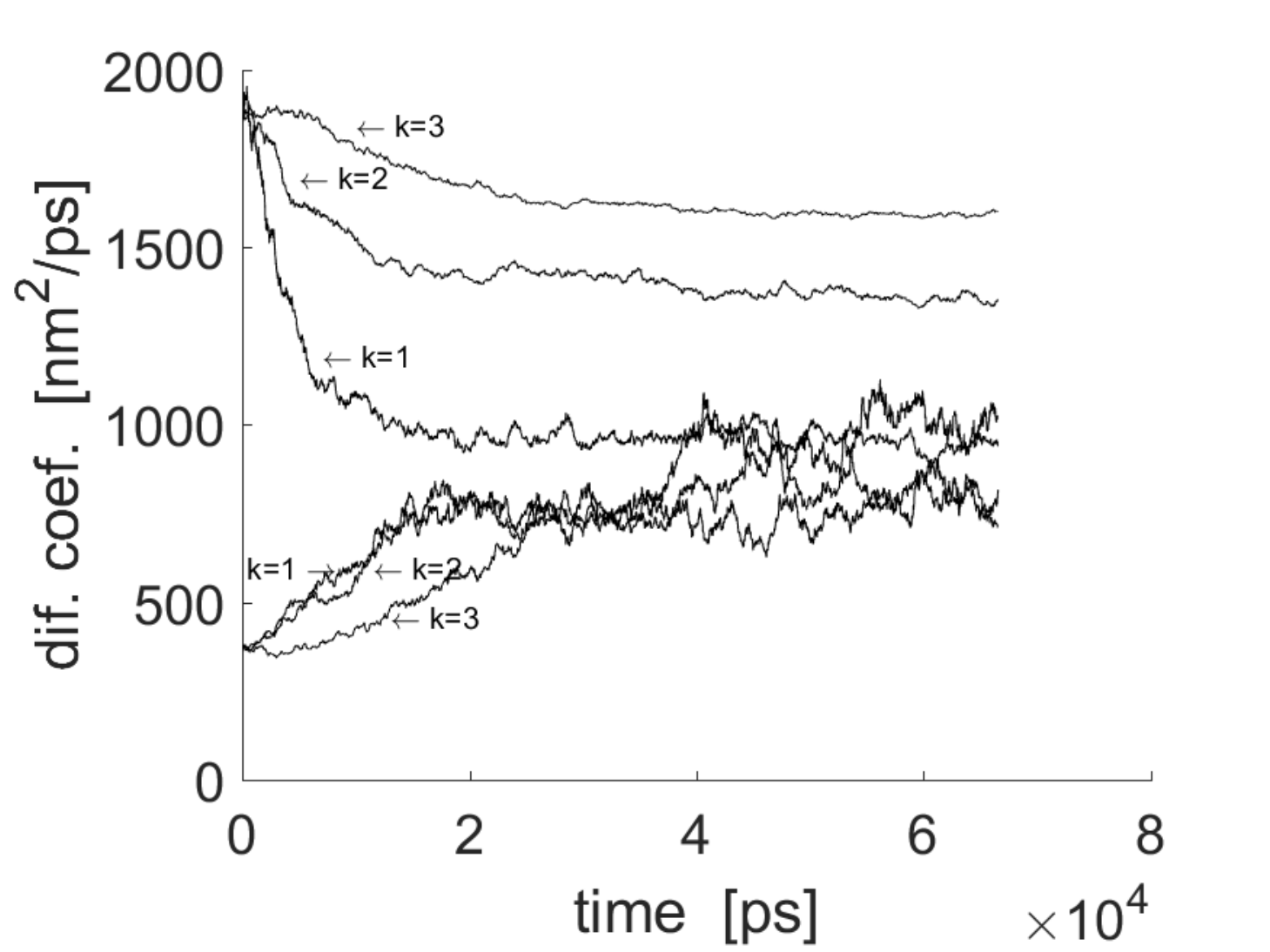}~
\includegraphics[totalheight=6cm,angle=0,keepaspectratio]{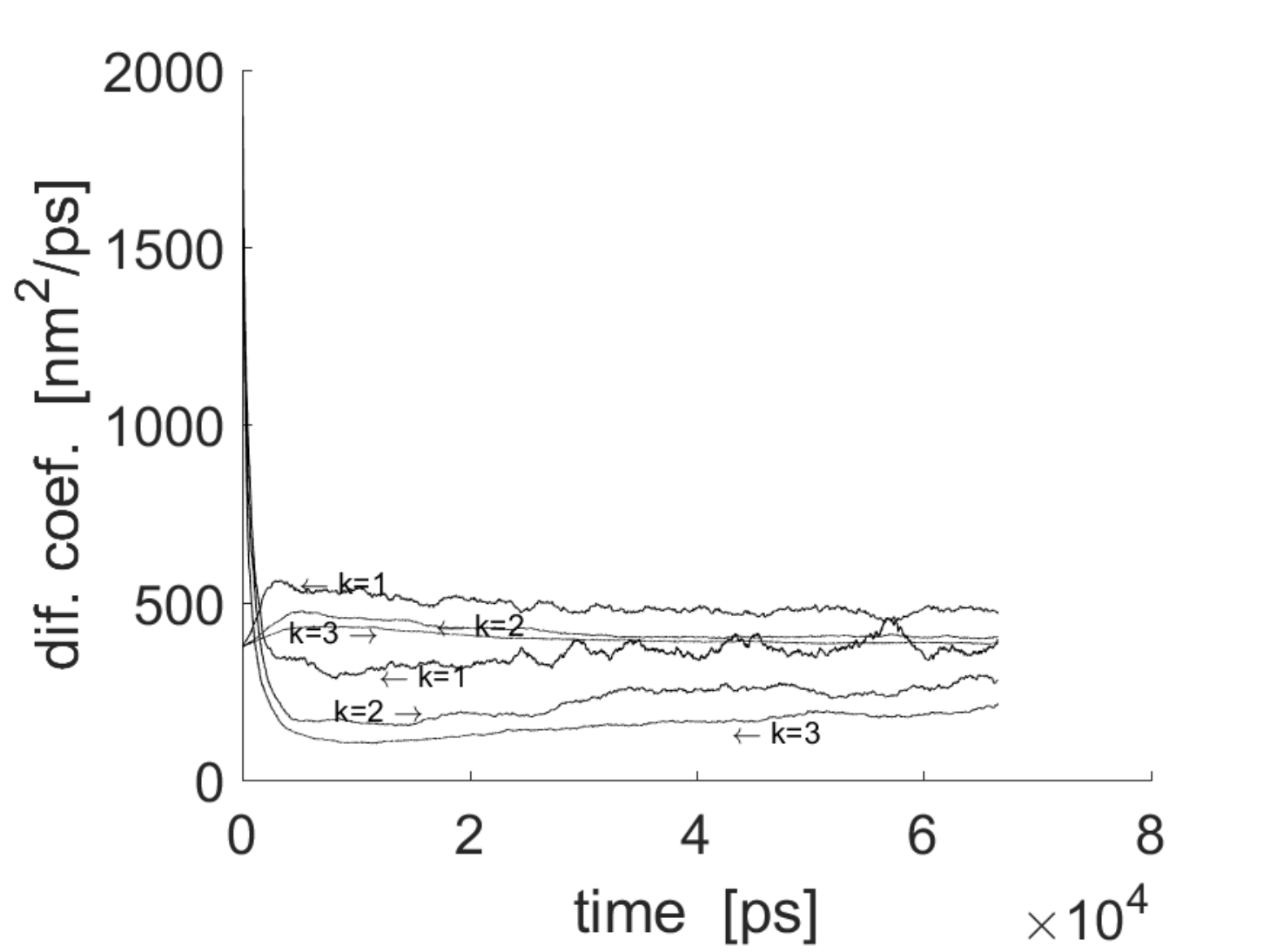}
\end{center}
\caption{Diffusion coefficients for drying process of the pore (left) and for wetting process of the pore  (right) for expanded outer spaces $k=1,2,3$ and final time $t_0=66532$ psec.}\label{Fig_k}
\end{figure}

The Figure \ref{Fig_k} refers to the case when the outer space is expanding $k=1,2,3.$ The Figure \ref{Fig_k} (left) concerns to the drying process while the Figure \ref{Fig_k} (right) concerns to the wetting process and both correspond to the same temperature, pressure and number of molecules like Fig. \ref{Fig_Dif_coef}. In the Fig. \ref{Fig_k} (left), upper curves starting near the point $(t,D)=(0,1900)$ correspond to the diffusion coeffients for the pore while the lower curves starting near the point $(t,D)=(0,400)$ correspond to the diffusion coeffients for the outer spaces. Similarly, in the Fig. \ref{Fig_k} (right), upper curves starting near the point $(t,D)=(0,400)$ correspond to the diffusion coeffients for the pore while the lower curves starting near the point $(t,D)=(0,1900)$ correspond to the diffusion coeffients for the outer spaces.       

\begin{figure}[H]
\begin{center}
\includegraphics[totalheight=6.0cm,keepaspectratio]{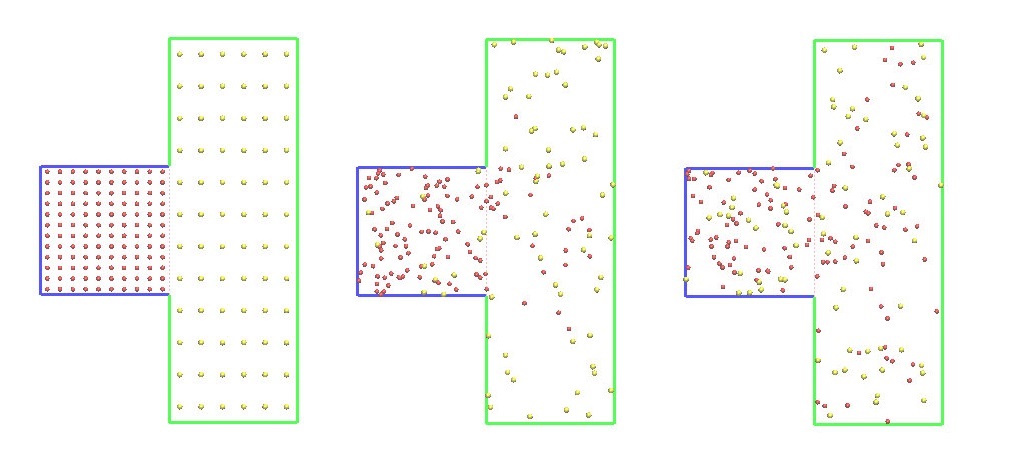} 
\end{center}

\caption{Water vapor molecules distributions in pore and outer space at the beginning (left) and at two time moments.}\label{Fig10}
\end{figure}

A visualization of one possible case of molecular dynamic evolution of water vapor molecules distribution in a pore and an outer space for three time steps are shown in the Figure \ref{Fig10}.

\paragraph{Computer simulation of macro-model}\label{macrosim}
In previous part, we showed the results of the micro-model for the case of the saturated water vapor at temperature $T_0=25$ $^oC$ and at pressure $p_0=3.17$ $kPa$ and the outer space $k=2$ (see Fig. \ref{fig2}). Beside this case, we calculated also two other cases when the saturated water vapor at temperature $T_0=15$ $^oC$ and at pressure $p_0=1.7$ $kPa,$ and the saturated water vapor at temperature $T_0=35$ $^oC$ and at pressure $p_0=5.62$ $kPa$ and for different outer spaces $k=1,2,3.$ Each of these cases provides a constant diffusion coefficient calculated according the formula (\ref{eq06}) for different outer spaces $k=1,2,3.$ The corresponding constant diffusion coefficient values are summarized in the Table \ref{Tab1}.      
\begin{table}[!h]
\centering
    \caption{Constant diffusion coefficients for different temperatures calculated according to the formula (\ref{eq06}); previously commented constants are in bold.}     
\begin{tabular}{|r||r|r|r||r|r|r|}
\hline
 &\multicolumn{3}{c||}{Drying}&\multicolumn{3}{c|}{Wetting}\\
\cline{2-7}
 T\ [$^oC$] &$k=1$&$k=2$&$k=3$&$k=1$&$k=2$&$k=3$\\
\hline\hline
15&1542.39&1994.85&2091.06&760.64&579.73&501.83\\
\hline
25&867.11&$\bf{1121.58}$&1169.05&427.82&$\bf{326.88}$&282.44\\
\hline
35&511.37&655.71&674.51&255.74&193.44&168.95\\
\hline
\end{tabular}\label{Tab1}
\end{table}
 
\begin{figure}[t]
\begin{center}
\includegraphics[totalheight=6cm,angle=0,keepaspectratio]{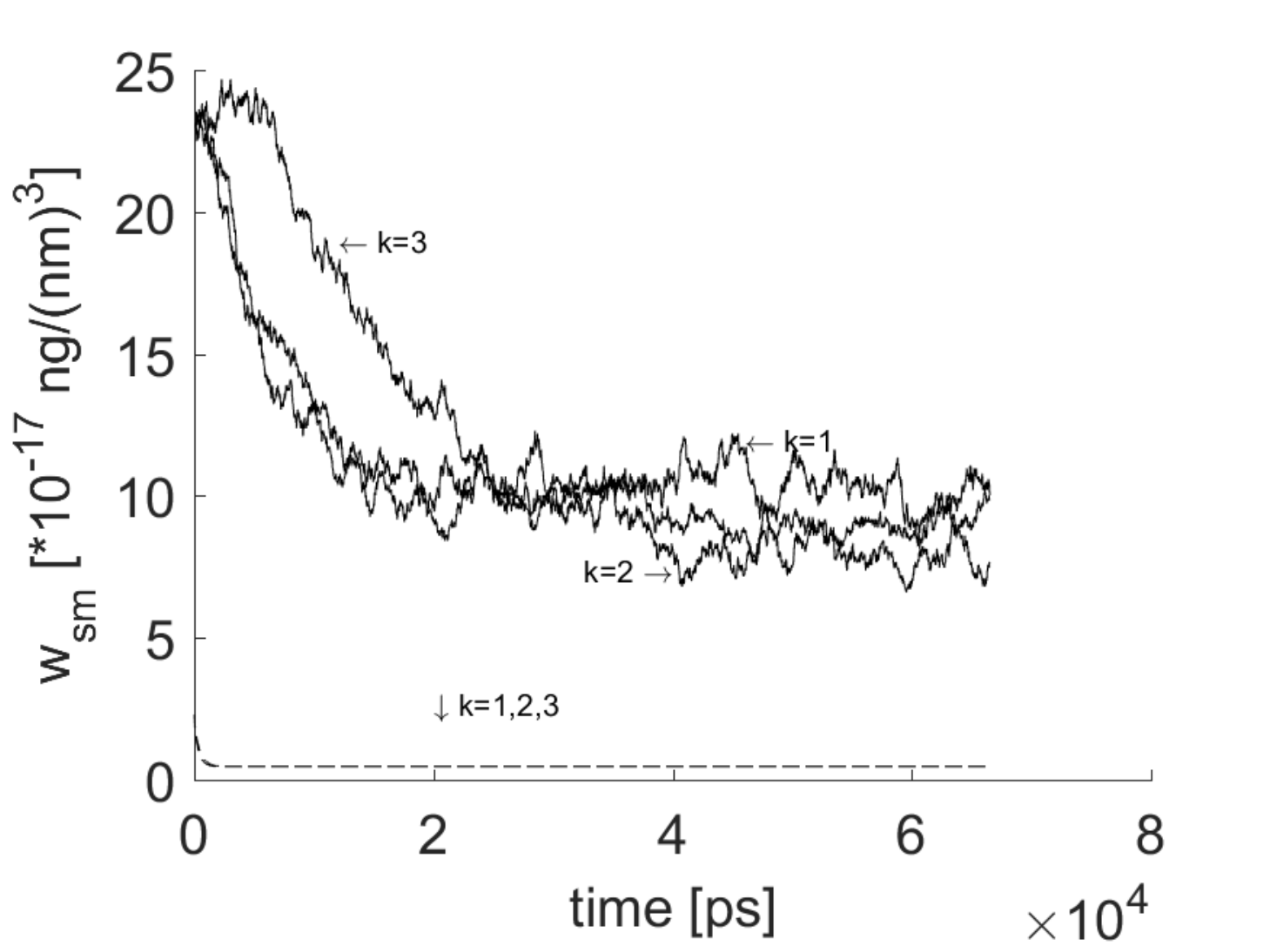}~
\includegraphics[totalheight=6cm,angle=0,keepaspectratio]{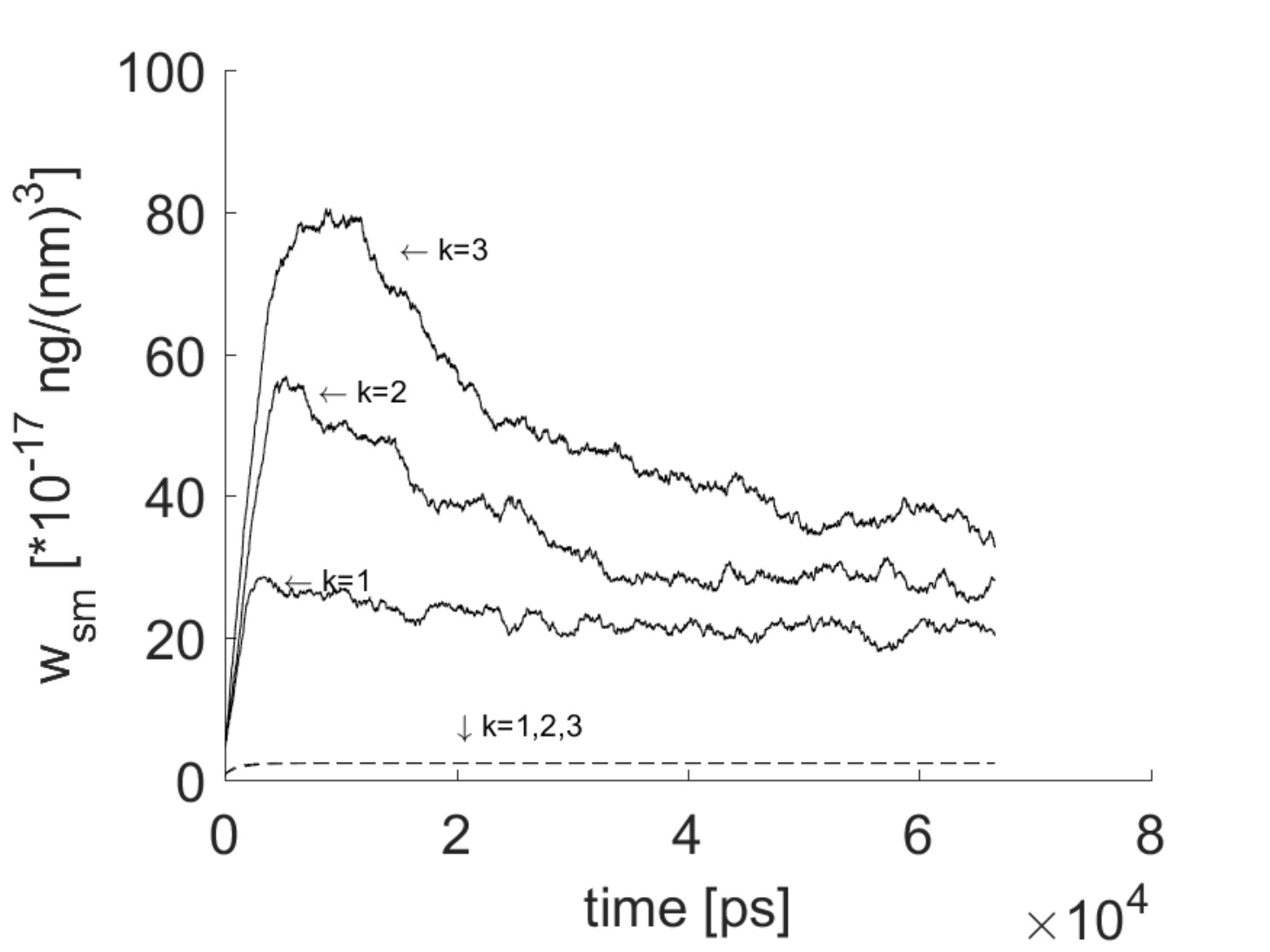}
\end{center}
\caption{Dynamics of the space mean of water vapor concentration (dashed curves) and dynamics of density (bulk curves) for final time $t_0=66532$  psec. Drying process (left) and wetting process (right).}\label{Fig6}
\end{figure}

The micro-model provides also another important characteristic like the density of $H_2O$ molecules in the pore. The density can be compared by a space mean of water vapor concentration which can be calculated from the solution (\ref{eq05}) according the following formula (\ref{eq07})
\begin{equation}\label{eq07}
w_{sm}(t)=\frac{1}{l_xl_y}\int_0^{l_x}\int_0^{l_y}w_v(x,y,t)dxdy
\end{equation}
Dynamics of the space mean of water vapor concentration $w_{sm}(t)$ and dynamics of density for final time $t_0=66532$ psec and for the saturated water vapor at temperature $T_0=25$ $^oC$ and at pressure $p_0=3.17$ $kPa$ and the outer space $k=2$ are shown in figure \ref{Fig6}.

\paragraph{Conclusions}
The both models give a consistent description of water vapor and a pore interaction in 2D case. The both models are consistent with experiment results concerning diffusion coefficient values. It is possible to use so called ''hybrid'' (a combination of micro and macro) approach for more accurate simulation of water vapor - pore interaction processes. But one can see from results of molecular dynamic simulations that the macro and the micro approaches are closer if a number of molecules used in molecular dynamic calculations is sufficient for selected ambient temperaure and pressure which agrees well with the conclusions of the article \cite{NorSte12}. It means that these calculations are also more realistic.
%

\end{document}